# On the rules of science game

Endel Põder

## Abstract

Credit allocation in the mainstream bibliometrics is fundamentally flawed and the popular indicators have been misleading science for decades. Originally a simple technical mistake has become an integral part of our culture and is very difficult to correct. Although the problem has been raised in scientific articles, it seems mostly unknown to wider audience.

Every year, Clarivate Analytics publishes a list of winners in the top science game – highly cited researchers.

During the last year, the number of highly cited researchers in Estonia dropped almost twofold – from 17 to 9. Someone may think that there is a problem with Estonian science. However, the real problem is very different. Analysts from Clarivate had noticed that, among highly cited authors, there are too many those who mostly publish with hundreds of coauthors. Admitting that this observation "strains their reason", they tried to reduce the number "bad guys".  However, instead of exclusion directly the papers with too long list of authors, the papers with more than 30 affiliations of the authors were excluded. Still, it helped to remove a part of "too heavily collaborating" people from the list of highly cited researchers.

Why should we worry about small chores of Clarivate Analytics? Because these are symptoms of a big problem of credit allocation in science.

Probably, many believe that position in the list of highly cited researchers is a good indicator of scientific achievements. I think it's rather problematic.

In order to be "highly cited", you need to be a coauthor of a maximum number of articles, with maximum number of cites per each. Cites per article is a widely accepted proxy for its quality and could be a measure of scientific achievement. Becoming a coauthor of a big number of articles is a different kind of achievement that needs social activity and skills of a salesperson. Because the final score is a multiplication of the two factors, both are equally important.

There is no evidence that the game was intended that way.

Citation analysis is based on two simple assumptions: 1) citation count is a reasonable measure of quality (or a "value") of an article, and 2) performance of a scientist can be measured by a total amount of the "value" he has produced, or sum of citations of the articles he has published. Implicitly it is assumed that each article has only one author. With several coauthors, we must somehow to divide the total amount of "value" between them. While coauthors may disagree on their relative contribution, there is no doubt that the sum of their contributions is exactly one article, with the "value" measured by its citation count.

Until we have no information on relative contributions, the best idea is to assume they are equal, and divide citations equally too.

The present whole-count system seems to be a thoughtless application of a single-author logic to multiple-authors case. It overlooks the difference between a coauthor and a full author and leads to a bizarre implication that every part is equal to the whole (1/2=1/5=1/10=1/100=1/1000=1). And this system was put into action.

Once the citation scores for individual researchers became available worldwide, they started to influence research and publication behavior. Some smart researchers noticed that working in a large group makes one a better scientist, in terms of publication and citation counts. Really, publishing articles together with a colleague, you can easily double your numbers of publications and citations as compared to publishing individually. And organizing a group of ten collaborators gives, in average, a tenfold advantage to all of them. Given such a strong incentive, it is not surprising that research groups are growing fast.

Nowadays, every scientist knows that collaboration is a key to success. Still, they are usually unaware about the biased bibliometrics behind this "law of nature". It is much simpler to believe that collective work really creates better science. And this may be even correct, sometimes. However, any real (positive or negative) effect on research efficiency is likely small compared to huge effects of collective publication.

Very likely, the pressure of bibliometric indicators has impact on who becomes a scientist, and which topics are being studied. A lonely genius cannot expect great success in the present conditions. Research problems that require a lot of fieldwork or gathering large data sets are preferred over these that need individual imagination primarily.

What can we expect in the future? The race of individual researchers for higher bibliometric scores will permanently push numbers of coauthors up, until a theoretical limit when all scientists are coauthors of all published articles. Perhaps at some point before that, authorship, and author-based bibliometrics, will lose any meaning and disappear.

Or is it possible that the scientific community will restart the game with correct rules? We have seen several calls for that during the last 40 years (Price, 1981; Schreiber, 2008; Põder, 2010). However, this possibility looks very unlikely. For sure, majority of influential people in science are the best players of the current game. Should they be happy to change the rules?